\documentclass{mem}
\usepackage{natbib}\usepackage{txfonts}\usepackage{balance}
\usepackage{graphicx}
\usepackage[a4paper,breaklinks,dvipdfm]{hyperref}
\idline{75}{282}
\begin{document}
\def\teff{$T\rm_{eff }$}
\def\kms{$\mathrm {km s}^{-1}$}
\def\Ms{M$_\odot$}
\title{``Dark'' systems in globular clusters: GWs emission and limits on the formation of IMBHs}

\subtitle{}

\author{M. \,Arca-Sedda\inst{1} }

\institute{University of Rome ``Sapienza'', Department of Physics,
P.le Aldo Moro, 5,
I-00185 Roma
\email{m.arcasedda@gmail.com}
}

\authorrunning{M. Arca-Sedda}
\titlerunning{''Dark'' systems in GCs}

\abstract{
Many observed globular clusters (GCs) seem to show a central overabundance of mass whose nature has not yet fully understood. Indeed, it is not clear whether it is due to a central intermediate mass black hole (IMBH) or to a massive stellar system (MSS) composed of mass segregated stars. In this contribution we present a semi-analytic approach to the problem complemented by 12 $N$-body simulations in which we followed the formation of MSSs in GCs with masses up to $3\times 10^5$ \Ms. Some implications for the formation of IMBHs and gravitational waves emission are discussed in perspective of a future work.
}

\maketitle{}

\section{Introduction}
The ever growing observational evidence of a mass excess in the centre of several globular clusters (GCs) \citep{noyola10}, suggests that they may contain at their centre an IMBH, likely formed through stellar runaway collisions \citep{zwart04}. On the other hand, 
these excesses may be due to massive stellar systems (MSSs) comprised of dark remnants of heavy stars which undergone mass segregation \citep{baumak03}. In this contribution we try to answer the question: ``can a MSS mimes efficiently the effects of an IMBH?'' 

To reach the aim, we developed a semi-analytic approach for modelling 168 star cluster models with different masses and global properties. Moreover, we used 12 detailed $N$-body simulations of GCs with masses up to $3\times 10^5$ \Ms. 
In order to estimate MSS sizes and masses, we followed the procedure described in \cite{AS16}.

\begin{figure*}[t!]
\resizebox{\hsize}{!}{
\includegraphics[clip=true]{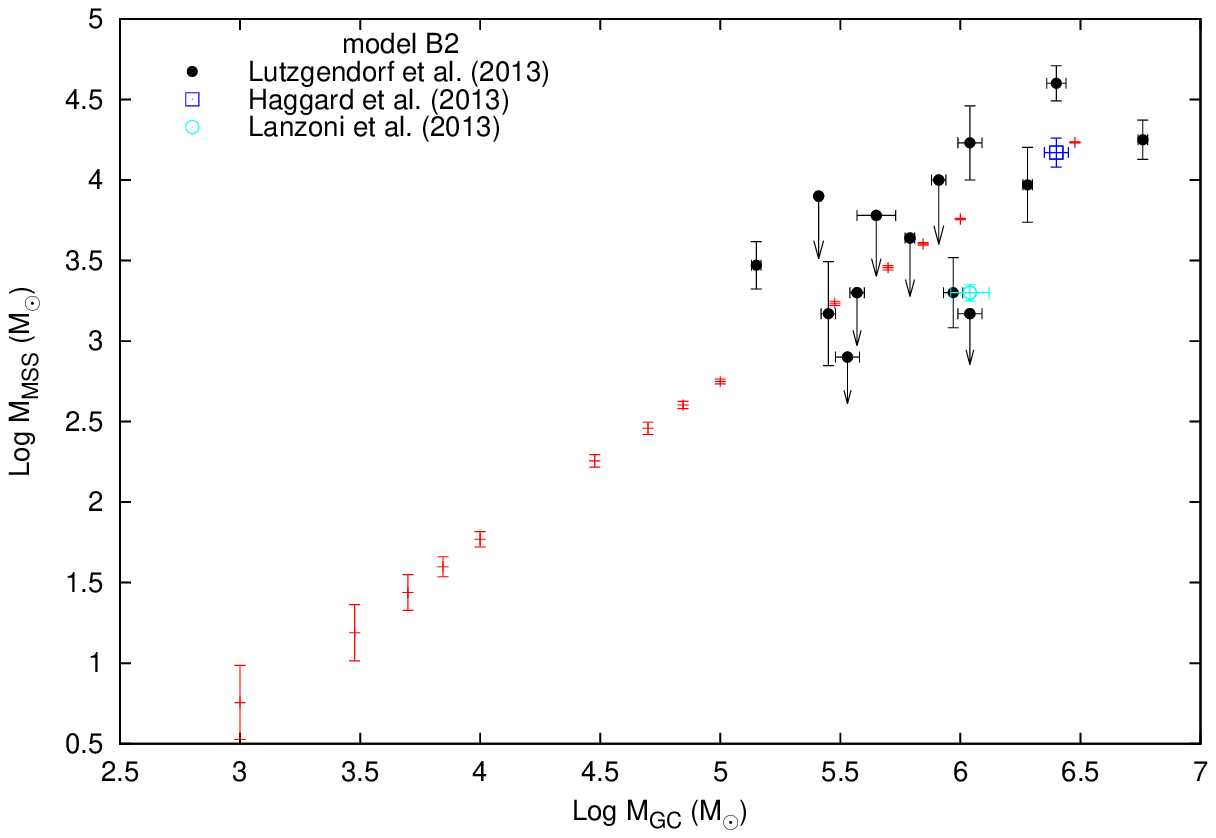}
\includegraphics[clip=true]{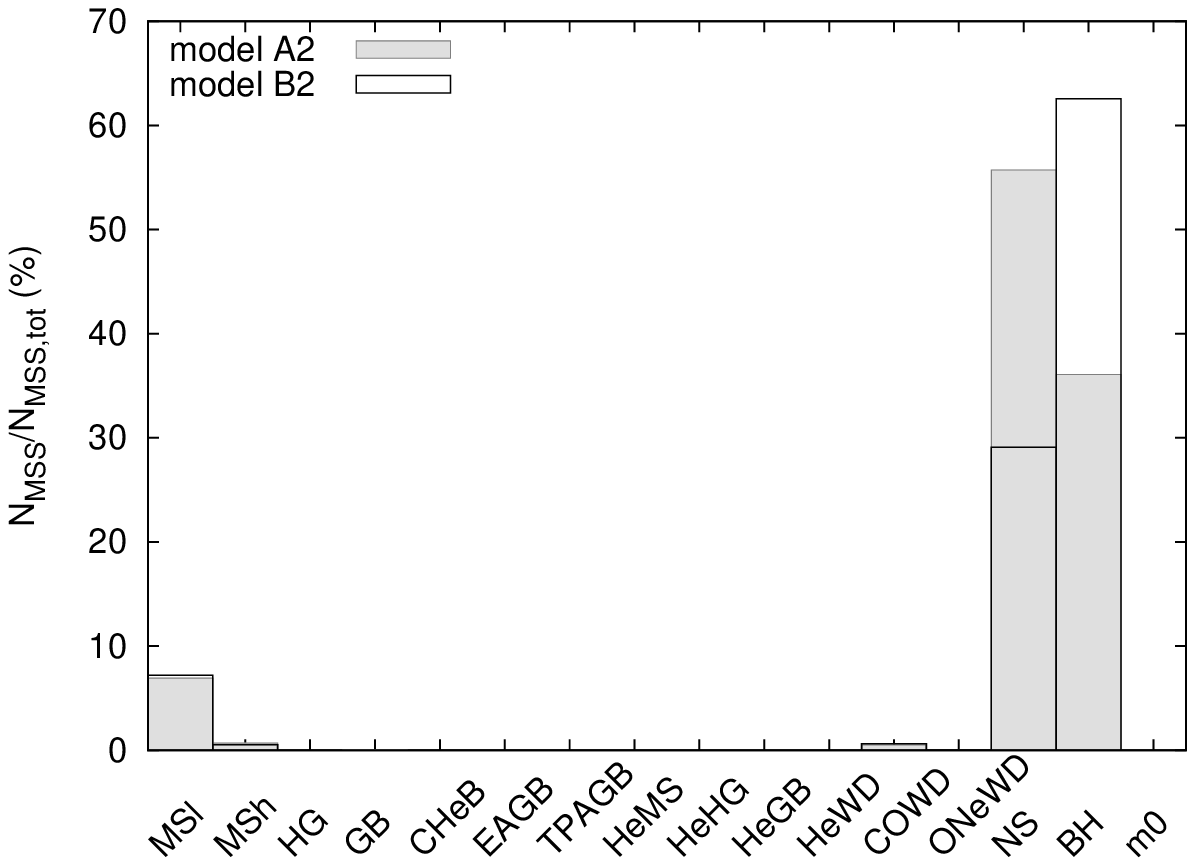}
\includegraphics[clip=true]{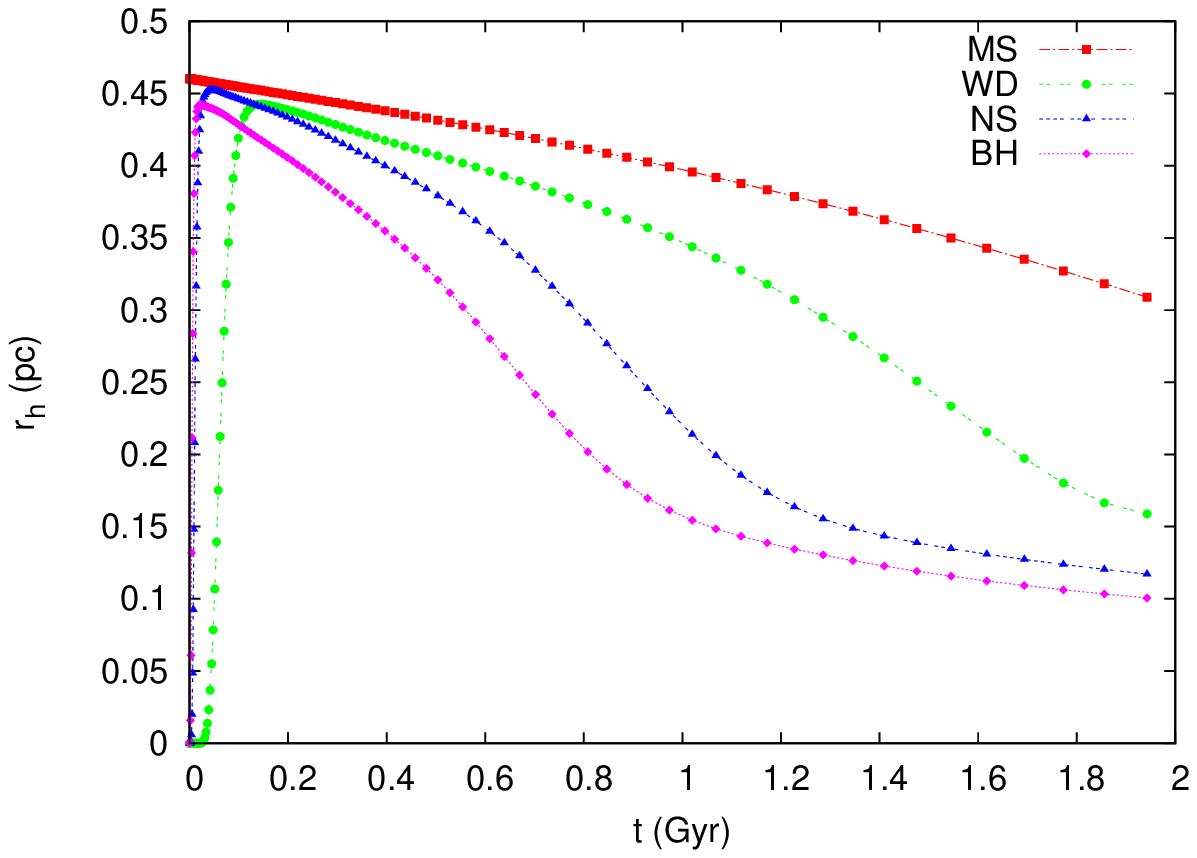}}
\caption{\footnotesize Left panel: MSS mass vs host cluster mass. The red dots with error bars represent our models, whereas black, blue and cyan symbols are observations. Central panel: number distribution of stellar kinds in a model with solar (model A2) and low (model B2) metallicity. Right panel: time evolution of the half-mass radii of different stellar types.}
\label{F1}
\end{figure*}

\section{Method and results}
To build our sample of models, we varied host GC masses ($10^3$-$3\times 10^6$ \Ms), initial mass function (IMF) (Kroupa, Salpeter and flat), density distribution and metallicity ($Z=0.02$-$0.0001$). To estimate mass segregation and stellar evolution, we used \texttt{CodeSSE}, a code which implements dynamical friction \citep{ASCD14df} and stellar evolution \citep{hurley}. 

We found that MSS masses correlates with the host cluster masses through a power-law, whose slope $\alpha$ depends on the cluster mass distribution. In all the considered cases, the value of $\alpha$ that best fits observations is $\sim 1$.
Interestingly, our results suggest that for low metal clusters the $60\%$ of their central MSS consist of stellar BHs, whereas for solar values of $Z$ such fraction decreases down to $20\%$. Moreover, stellar types are strongly mass segregated, with the BH population having a half-mass radius smaller than $0.1$ pc (see Fig. \ref{F1}).

Our semi-analytic results are well reproduced by $N$-body simulations, which we performed using the code \texttt{HiGPUs} \citep{Spera} and \texttt{HiGPUsSE} \citep{AS16}, using up to $524$ k particles. The velocity dispersion profiles for our model are similar to observations, possibly suggesting that a MSS can induce the same dynamical feedback of an IMBH.

Recently, we improved our semi-analytic treatment by including the package \texttt{BSE} \citep{hurley02} in our code, whose new version is named \texttt{CodeBSE}. Our preliminary results indicate that the MSS-host cluster correlations are poorly affected by primordial binaries as long as their fraction is $\lesssim 50\%$.

BHs mass segregation can lead to the formation of BH binaries (BHBs) and possibly, to merging events that can produce detectable bursts of gravitational waves, as recently observed by the LIGO/Virgo collaboration \citep{LIGO}. 

In this context, it is easy to show that the minimum mass for a GC to host at least 2 stars with masses $>120$ \Ms is $M_{\rm GC}>10^4$ \Ms, for a Kroupa IMF.
To investigate the possible formation of a massive BBH in such environment we developed several simulations with \texttt{ARWc-M}, a code which implements the algorithmic regularization chain \citep{mikkola99}. 
We found that in most cases a series of scatterings lead to the formation of a triple system, composed of a BH orbiting around a BBH. In one of the models, the BHB had semi-major axis $a=20$ a.u. and components with masses $M_1 = 13$ \Ms and $M_2=22$ \Ms, whereas the single BH had mass $M_3 = 25$ \Ms orbiting at $200$ a.u. from the BHB. After $10$ Myr, the BBH is kicked out. 
Following \cite{peters64}, the time-scale for GWs emission by the BBH in this case was $t_{\rm GW}=2\times 10^9$ Gyr. 

However, we discovered an interesting tidal effect originating when $M_3$ moves on the same-plane of the BBH but it counter rotates. In this configuration, a very fast interaction occurs over time-scales $< 10^4$ yr and kicks out the BBH, whose GWs time-scale becomes $t_{\rm GW}\lesssim 1$ Gyr. 

In a forthcoming work, we will try to determine the occurrence of these effects in young massive clusters.

%\begin{figure*}[t!]
%\resizebox{\hsize}{!}{\includegraphics[clip=true]{}}
%\caption{\footnotesize
%
%}
%\label{}
%\end{figure*}

%\begin{table*}
%\caption{}
%\label{}
%\begin{center}
%\begin{tabular}{}
%\hline
%\\
%\hline
%\end{tabular}
%\end{center}
%\end{table*}

\begin{acknowledgements}
MAS acknowledges financial support from the University of
Rome ``Sapienza'' through the grant ``52/2015'' in the frame-
work of the research project ``MEGaN: modelling the envi-
ronment of galactic nuclei''
\end{acknowledgements}

\footnotesize{
\bibliographystyle{aa}
\bibliography{bblgrphy}
}

\end{document}